\documentclass{elsart}
\usepackage{graphicx,harvard,amssymb}

\makeatletter
\def\elsartstyle{%
        \def\normalsize{\@setfontsize\normalsize\@xiipt{14.5}}
        \def\small{\@setfontsize\small\@xipt{13.6}}
        \let\footnotesize=\small
        \def\large{\@setfontsize\large\@xivpt{18}}
        \def\Large{\@setfontsize\Large\@xviipt{22}}
        \skip\@mpfootins = 18\p@ \@plus 2\p@
        \normalsize
}
\makeatother

\def\url#1{{\ttfamily\def\/{/\discretionary{}{}{}}#1}}

\def\gte{\,\lower.6ex\hbox{$\buildrel >\over \sim$} \, }
\def\lte{\,\lower.6ex\hbox{$\buildrel <\over \sim$} \, }

\begin{document}

\begin{frontmatter}
\title{Microlensing of Extended Stellar Sources}

\author[Glasgow]{M.A. Hendry\thanksref{email}}
\author[Glasgow]{I.J. Coleman\thanksref{current}}
\author[Glasgow]{N. Gray}
\author[Soton]{A.M. Newsam}
\author[Glasgow]{J.F.L. Simmons}

\address[Glasgow]{Department of Physics and Astronomy, University of Glasgow, Glasgow UK}
\address[Soton]{Department of Physics and Astronomy, University of Southampton,\\
Southampton UK}

\thanks[email]{E-mail: martin@astro.gla.ac.uk}
\thanks[current]{Current address: British Antarctic Survey, Cambridge, CB3 0ET, UK}

\begin{abstract}
We investigate the feasibility of reconstructing the radial
intensity profile of extended stellar sources by inverting their
microlensed light curves. Using a simple, linear, limb darkening
law as an illustration, we show that the intensity profile can be
accurately determined, at least over the outer part of the stellar
disc, with realistic light curve sampling and photometric errors.
The principal requirement is that the impact parameter of the lens
be less than or equal to the stellar radius. Thus, the analysis of
microlensing events provides a powerful method for testing stellar
atmosphere models.
\end{abstract}

\end{frontmatter}

\section{Introduction}
\label{intro}

In recent years a number of authors have investigated the
microlensing of extended stellar sources. \citeasnoun{SNW95}
(hereafter SNW95) have shown that the light curves of extended
sources can exhibit a significant chromatic dependence, essentially
because limb darkening renders the effective radius of the star a
function of wavelength. Thus, in addition to improving constraints
on the lens parameters, modelling the microlensing of extended
sources provides a powerful tool for gravitational imaging stellar
surfaces.

In  this contribution we describe how microlensing light curves of
extended sources may be used as a test of stellar atmosphere
models. We generate artificial light curves, assuming a particular
model atmosphere, and use the Backus-Gilbert numerical inversion
method to estimate the radial stellar intensity profile from the
observed light curves.

\section{Inverting Microlensing Light Curves}
\label{invert}

The (time dependent) integrated flux, $F(t)$, from an extended
stellar source of radius, $R$, lensed by a point lens is given by
(c.f. SNW95)
\begin{equation}
F(t) = \int_{s=0}^{R} \int_{\theta=0}^{2\pi} I(s,\theta) A(d) s ds
d\theta
\end{equation}
where $d$ is the projected distance from the lens to the element of
the stellar surface and the amplification function, $A(d)$, takes
its usual analytic form for a point source. (Note that $d$ is a
function of $s$, $\theta$ and $t$).

If we assume that the projected stellar surface displays circular
symmetry -- i.e. $I(s,\theta) \equiv I(s)$ -- we may write the
above equation in the form
\begin{equation}
F(t) = \int_{r=0}^{1} I(r) K(r,t) ds
\end{equation}
where $r = s/R$ and the kernel function, $K(r,t)$, is obtained by
integrating over $\theta$. A solution to this integral equation for
$I(r)$ can be obtained by applying the Backus-Gilbert inversion
procedure. This method takes account of the smoothing effect of the
kernel function and reconstructs a regularised estimator of $I(r)$
which optimises the trade-off between the bias and variance of the
estimator. For details of the method see \citeasnoun{CGS98} and
\citeasnoun{C98}.

\section{Testing Stellar Atmosphere Models}
\label{atmos}

In order to test the feasibility of reconstructing stellar surface
profiles, we assumed a simple, linear, limb darkening law with
coefficient, $u$. We generated microlensed light curves of
typically 100 data points, with an even sampling rate, and Gaussian
noise added to the photometry at a level of typically 2\% of the
baseline flux. In most cases we took the impact parameter and the
einstein radius equal to the stellar radius. We considered two
model atmospheres:
\begin{center}
\begin{trivlist}
\item[(1)]{$T_{\rm{eff}} = 4000 K$; $\log g = 2.0$; $u_B = 0.92$, $u_V = 0.82$}
\item[(2)]{$T_{\rm{eff}} = 6730 K$; $\log g = 4.5$; $u_B = 0.76$, $u_V = 0.63$}
\end{trivlist}
\end{center}
where the Johnson $B$ and $V$ band linear limb darkening
coefficients are from \citeasnoun{CG90}. We carried out inversions
of the $B$ and $V$ band intensity profiles for these models, and
investigated the effect of changing the impact parameter, stellar
radius, light curve sampling rate and photometric accuracy.

\section{Results}
\label{results}

\begin{figure}
\begin{center}
\includegraphics*[width=6cm,angle=-90]{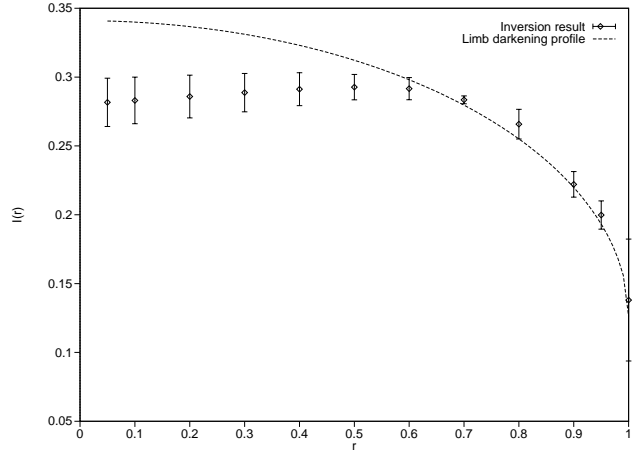}
\end{center}
\caption{Recovered radial $V$ band intensity profile from a light curve
of 100 points. The true profile, for $u_V
= 0.63$, is shown by the dotted line}
\label{fig:fig1}
\end{figure}

Figure \ref{fig:fig1} illustrates the reconstructed $V$ band
profile for model (2), for the case of the impact parameter and
stellar radius equal to the einstein radius of the lens. The errors
on the recovered solution are determined from the covariance matrix
of the Backus-Gilbert estimator. We can see that the reconstructed
profile is significantly biased for $r \lte 0.6$, due to the
smoothing effect of the kernel function, but the true profile is
well recovered over the interval $0.6 \lte r \lte 0.9$. We now
briefly summarise the results of varying the stellar and lens
parameters.

\begin{itemize}
\item{The inversions generally recover the true profiles well over the
interval $0.6 \lte r \lte 0.9$, for a wide range of $T_{\rm{eff}}$,
$\log g$ and $u$.}
\item{Increasing the einstein radius improves significantly the inversions for
$r \gte 0.6$, but no improvement is seen for $r \lte 0.6$.}
\item{Reducing the impact parameter (i.e. a transit event) significantly
improves the accuracy of the reconstruction for $r \gte 0.6$ and
reduces the bias for $r \lte 0.6$. For impact parameters greater
than the stellar radius, however, the reconstruction deteriorates
rapidly for all $r$.}
\item{Even with photometric errors of 10\% a reasonable recovery of $I(r)$
is still obtained over the interval $0.6 \lte r \lte 0.9$; on the
other hand, reducing the errors to only 0.2\% does {\em not}
improve the recovery for $r \lte 0.6$, however. This is because the
bias is primarily due to the ill-posedness of the kernel over this
range, and not due to the photometric errors.}
\item{The reconstructions become unacceptably noisy when the number
of light curve data points is reduced to $n = 10$, but there is
little further improvement in accuracy above $n=50$.}
\end{itemize}

\section{Conclusions}
\label{conc}

Our results indicate that -- with realistic light curve sampling
and photometric errors -- one can accurately reconstruct, at least
in part, the multicolour radial intensity profiles of extended
stellar sources from their microlensed light curves, provided that
the impact parameter of the lens is comparable to the stellar
radius. The smoothing properties of the kernel function result in a
biased solution for $r \lte 0.6$, unless the lensing event is a
transit with small impact parameter. Nevertheless, the accurate
recovery over the interval $0.6 \lte r \lte 0.9$ is a robust result
over a wide range of stellar temperatures and limb darkening
coefficients. Despite the narrow width of this `good fit' annulus,
it is still adequate to usefully discriminate between different
model atmospheres -- e.g. two models with the same temperature but
with $\log g = 2.0$ and $\log g = 4.0$. Thus, we conclude that
broad band microlensed photometric light curves are a powerful tool
for investigating extended stellar sources and testing model
stellar atmospheres, and form a useful adjoint to spectroscopic and
polarimetric microlensing signatures. We are currently
investigating the application of inversion techniques to more
realistic model atmospheres and stellar intensity profiles.

\end{document}